# Climate Change: Sources of Warming in the Late 20th Century

*Gerald E. Marsh*

Argonne National Laboratory (Ret)
5433 East View Park
Chicago, IL 60615

E-mail: gemarsh@uchicago.edu

**Abstract.** The role of the North Atlantic Oscillation, the Pacific Decadal Oscillation, volcanic and other aerosols, as well as the extraordinary solar activity of the late 20th century are discussed in the context of the warming since the mid-1970s. Much of that warming is found to be due to natural causes.

**INTRODUCTION**

In 1996 James Hurrell—with reference to the 1995 report of the Intergovernmental Panel on Climate Change—made the point that "elements of the temperature anomaly pattern since the mid-1970s resemble the greenhouse warming fingerprint predicted by some general circulation models. However, it is difficult to assess whether the observed changes are in response to greenhouse gas forcing, or whether the changes are part of a natural decadal-timescale variation in the circulation." He found that "Pronounced changes in the wintertime atmospheric circulation have occurred since the mid-1970s over the ocean basins of the northern Hemisphere, and these changes are related to changes in the north Atlantic Oscillation (NAO) . . ." and that "nearly all the cooling in the northwest Atlantic and the warming across Europe and downstream over Eurasia since the mid-1970s results from the changes in the NAO, and the NAO accounts for 31% of the hemispheric interannual variance over the past 60 winters."[1]

Later, in 2001, Visbeck, Hurrell, Polvani, and Cullen wrote that "Because global average temperatures are dominated by temperature variability over the northern land masses, a significant fraction of the recent warming trend in global surface temperatures can be explained as a response to observed changes in atmospheric circulation. Because the NAO is a natural mode of the atmosphere, one could argue that much of the recent warming is not related to the build-up of greenhouse gases in the atmosphere over the past century. This viewpoint, however, ignores the possibility that anthropogenic climate change might influence modes of natural variability, perhaps making it more likely that one phase of the NAO is preferred over the other." They went on to note, "At present, there is no consensus on the process or processes that are responsible for observed low-frequency variations in the NAO."[2]

Put in other terms, the buildup of atmospheric concentrations of carbon dioxide may not be directly responsible for most of the global temperature rise since the mid-1970s. Rather, forcing due to increased carbon dioxide concentrations or other human activities *may* have affected the duration and frequency of positive NAO phases.

Recent work in 2009 by Trouet, et al.[3] has shown that the Medieval Climate Anomaly (MCA), also known as the Medieval Warm Period, was the result of a persistent positive NAO. Specifically, [4] "The persistent positive phase reconstructed for the MCA appears to be associated with prevailing La Niña-like conditions possibly initiated by enhanced solar irradiance and/or reduced volcanic activity and amplified and prolonged by enhanced AMOC", where AMOC stands for Atlantic meridional overturning circulation.

**CLIMATE IMPACT OF VOLCANIC ACTIVITY**

While the volcanic record does show somewhat reduced activity during the Medieval Warm Period, reduced volcanic activity is not likely to be the dominant factor for changes in the NAO. For one thing, volcanic eruptions generally only have an impact on climate for a period of about six months, although some, such as Tambora in 1815, can affect climate for a couple of years—the year of 1816 was known as "the year without a summer" in New England and Canada. Periods of reduced volcanic activity, as will be seen, also coincide with strong negative phase NAOs. If *decreased* volcanic activity is to be causally related to positive phase NAOs, *increased* volcanic activity should also correlate with negative phase NAOs or the termination of positive phase NAOs. The volcanic record does not show this to be the case.

To have a one to two year effect on climate, sulphate from volcanic eruptions must enter the stratosphere since material injected into the troposphere is rapidly precipitated. In addition, the particles comprising the stratospheric sulfate aerosol must be of the appropriate size to induce a net cooling.

The volcanism record is generally based on electrical conductivity or sulphate measurements in ice cores. The sulphate record so obtained, however, must be adjusted for ocean productivity. Marine phytoplankton (mostly algae) produce dimethylsulfide ($C_2H_6S$). More extensive winter sea ice promotes an increase in phytoplankton activity during the seasonal melting of sea ice, resulting in an increase in the amount of dimethylsulfide released.

The principal oxidation products of dimethylsulfide are methanesulfonic acid ($CH_3SO_3H$) and sulfur dioxide ($SO_2$), which is converted to non-sea-salt sulphate. The latter is also produced by volcanoes and other sources of non-organic sulphate. It has been maintained that methanesulfonic acid in Antarctic ice cores can be used as a measure of marine biogenic activity[5] or sea ice extent.[6] Unfortunately, the oxidation reactions that convert dimethylsulfide to methanesulfonic acid and sulfur dioxide are not well known, and the relative amounts of these products—and variations in their transport to Antarctica—are uncertain.[5]

Even if one could convincingly correct the ice core record for ocean productivity, there is no direct way to convert ice core measurements into global radiative forcing. Sulfur dioxide injected into the stratosphere is converted into a sulfate aerosol by a series of reactions similar to

$$SO_{2(gas)} + OH \rightarrow HOSO_2$$
$$HOSO_2 + O_2 \rightarrow SO_3 + HO_2$$
$$SO_3 + H_2O \rightarrow H_2SO_{4(liq)}$$

The hydroxyl radical (OH) is probably the most important oxidant. Following large eruptions, the scattering of solar radiation by the aerosol and reaction with $SO_2$ can deplete the abundance of OH radicals in the stratosphere thereby affecting the rate of conversion of sulfur dioxide to aerosol.

At the same time, the material injected into the stratosphere begins to circle the globe equatorially and is also unevenly transported latitudinally. In addition, condensation and coagulation produce larger particles that do not scatter incoming radiation as effectively as smaller ones and also settle out of the stratosphere at a faster rate.

Stratospheric aerosols have a cooling effect by reflecting short-wave solar radiation,

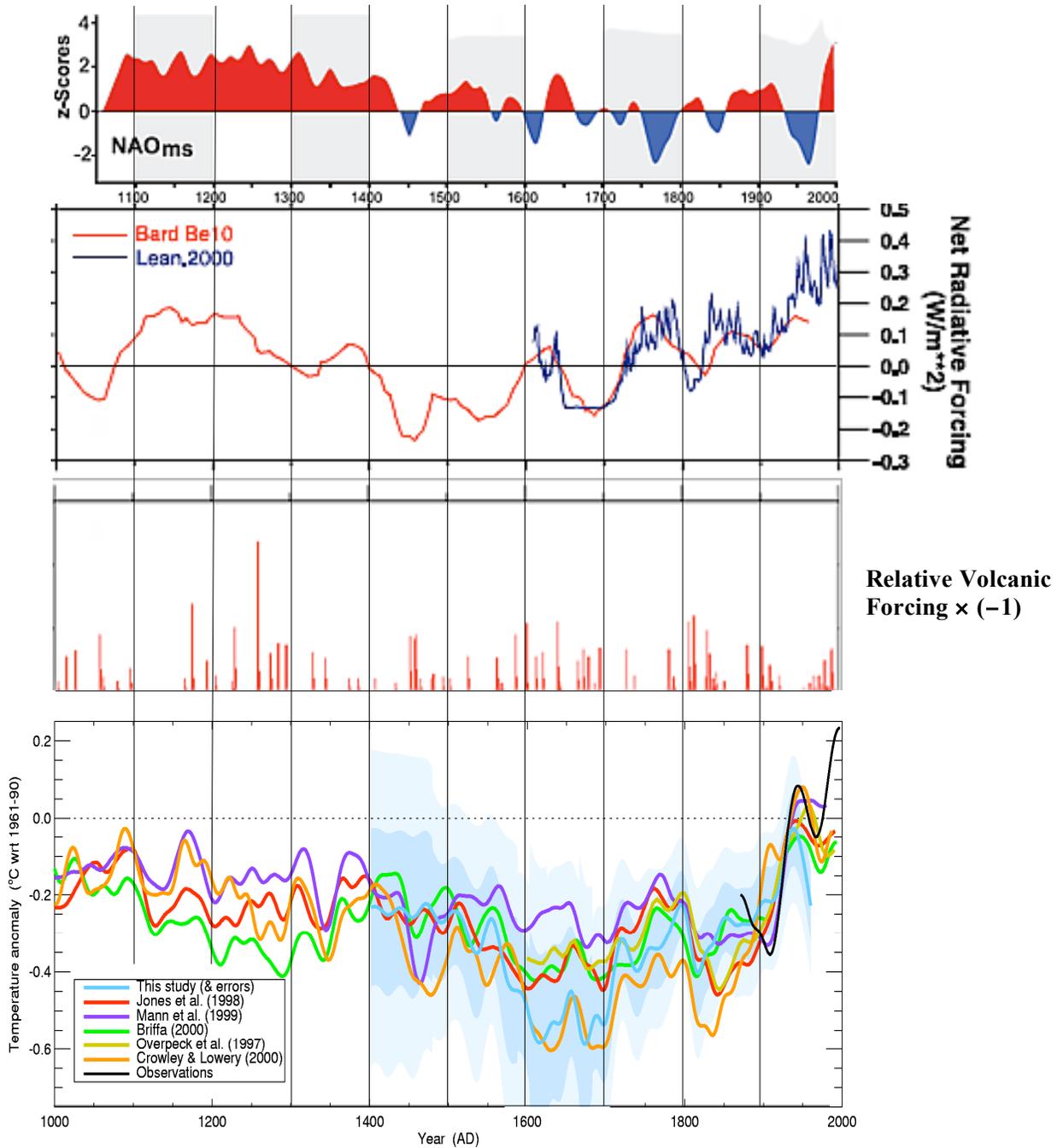

Fig. 1. The NAO record as adapted from Trouet, et al., and the solar and volcanic records as adapted from Crowley. The time line is the same for the four figures. *z*-score is a measure of the number of standard deviations from the mean. Trouet, et al. used a 30-year spline smoothing with the normalization period of 1500-1983. The temperature record compares six large-scale reconstructions, all recalibrated with linear regression against the 1881-1960 mean April-September observed temperature averaged over land areas north of 20N. All series have been smoothed with a 50-year Gaussian-weighted filter and are anomalies from 1961-1990 mean (from K. R. Briffa, et al.[7]).

thereby increasing the albedo of the earth. Aerosols can also have a warming effect by reflecting infrared radiation back to the earth. Which of these two effects dominates depends on the size and shape of the aerosol particles. For spherical particles, the critical radius is ~2 $\mu$m. Particles having a radius smaller than this value cause cooling. Thus, while the above considerations imply that the radiative effects of volcanism are difficult to estimate historically and are likely to be self-limiting,[8] for large eruptions global radiative cooling can certainly be significant for a period of one to two years. The Pinatubo eruption in June of 1991, for example, led to a radiative cooling of ~3 $w/m^2$ by October 1991. Sparks, et al, have discussed the atmospheric effects of the Pinatubo eruption in detail.[9]

The relationship between the NAO and the solar, temperature, and volcanic record is shown in Fig. 1. Although radiative forcing is given by Crowley[10] in the volcanic record shown in this figure, the estimated values are not reproduced here because of the uncertainties given above. What is important for the purposes of this essay is the time of the eruptions in the context of the NAO record.

With regard to the possibility that reduced volcanic activity could be responsible for inducing a positive phase in the NAO, note that the deepest *negative* phase NAOs occur from ~1750-1800 and ~1930-1975. Both these periods correspond to intervals of reduced volcanic activity.

As mentioned earlier, if decreased volcanic activity is to be causally related to positive phase NAOs, increased volcanic activity should correlate with negative phase NAOs or the termination of positive phase NAOs. The largest eruption in terms of relative forcing seen in the volcanic time series of Fig. 1 was in 1259. It had little or no effect on the extended Medieval Warm Period NAO and in fact occurred just before a slight positive increase in the NAO *z*-score. The 1815 Tambora eruption that caused the "year without summer" also did not impede the rise of a positive NAO. Krakatoa in 1883 did not terminate an existing positive NAO, although one can see a slight decrease in this positive NAO following the eruption. However, this positive NAO, if it was indeed affected by Krakatoa, recovered quickly and continued to rise until about 1920 when its *z*-score began to decrease. Note also that the 1902 Santa Maria eruption did not prevent the continued rise of a positive NAO. This is also true of the 1991 Pinatubo eruption.

From Fig. 1, it would seem that the correlation between negative phases of the NAO and volcanic activity, and the absence of volcanic activity with positive phases, is rather weak at best. On the other hand, the correlation between the NAO and the temperature record and solar radiative forcing is much better over the entire 1100-2000 period.

The Visbeck, et at. argument that "anthropogenic climate change might influence modes of natural variability, perhaps making it more likely that one phase of the NAO is preferred over the other" cannot be decided by correlation over the limited period available. Most of the anthropogenic carbon dioxide was put into the atmosphere after ~1940. The period from ~1940 to 1976-1977 was dominated by a large negative NAO—see Fig. 1—followed by a large positive NAO. Since similar positive NAOs have occurred in the past, it cannot be said that the latest is due to human activity simply because it correlates with rising carbon dioxide concentrations.

It is interesting that the large negative NAO that began in the earlier portion of the $20^{th}$ century and extended to about 1975 roughly corresponds to a negative phase of the Pacific Decadal Oscillation (PDO) shown in Fig. 2. And the recent large positive NAO also corresponds to the positive shift of the PDO of 1976-1977.

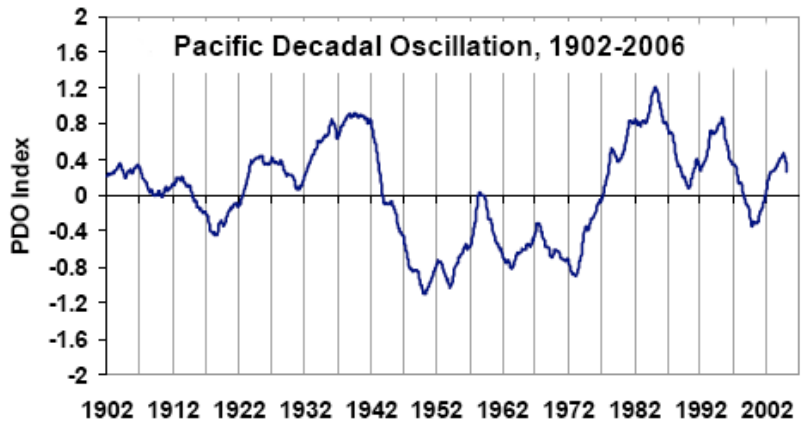

Figure 2. The Pacific Decadal Oscillation from 1902-2006. [Adapted from a presentation by R. W. Spencer, "Global Warming as a Response to the Pacific Decadal Oscillation" ,The University of Alabama in Huntsville (15 December, 2008)]

However, the correlation does not appear to be robust when compared to PDOs extending back to 1600, as seen in Figure 3. In terms of the temperature shift in the arctic, however, the impact of the positive phase of the PDO—often called "The Great Pacific Climate Shift of 1976-1977"—has been dramatic. Composite temperatures from Fairbank, Anchorage, Nome and Barrow (see Fig. 4) show a rise of ~1.4 °C followed by a *decrease* of ~0.24 °C/decade.[11]

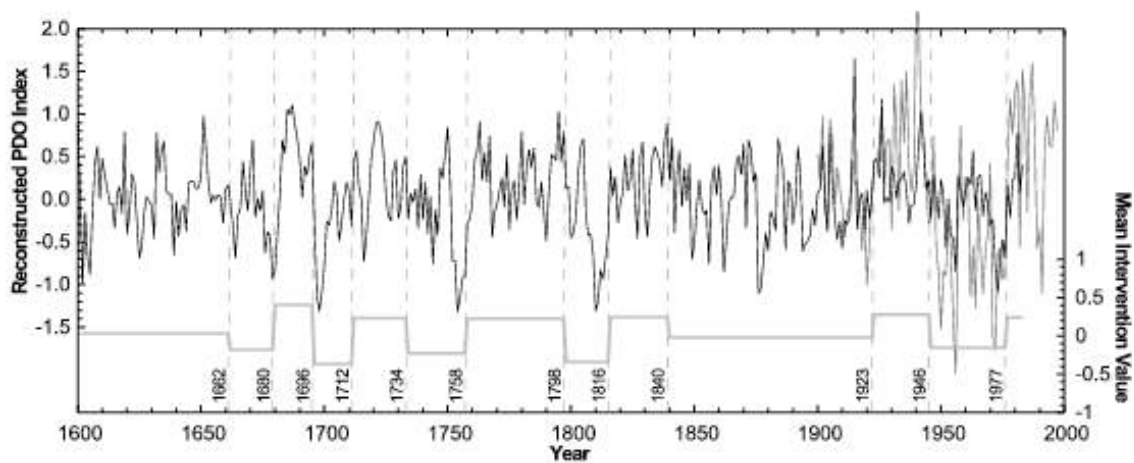

Figure 3. Tree ring-width chronologies from coastal western North America as a proxy for the PDO since 1600. Observed values are shown by the grey curve and reconstructed ones by the black. The bottom curve gives shift in the PDO using an "intervention model". [From Z. Gedalof and D. J. Smith, "Interdecadal climate variability and regime-scale shifts in the Pacific North America", *Geophys. Res. Lett.* **28**, 1515-1518 (2001).]

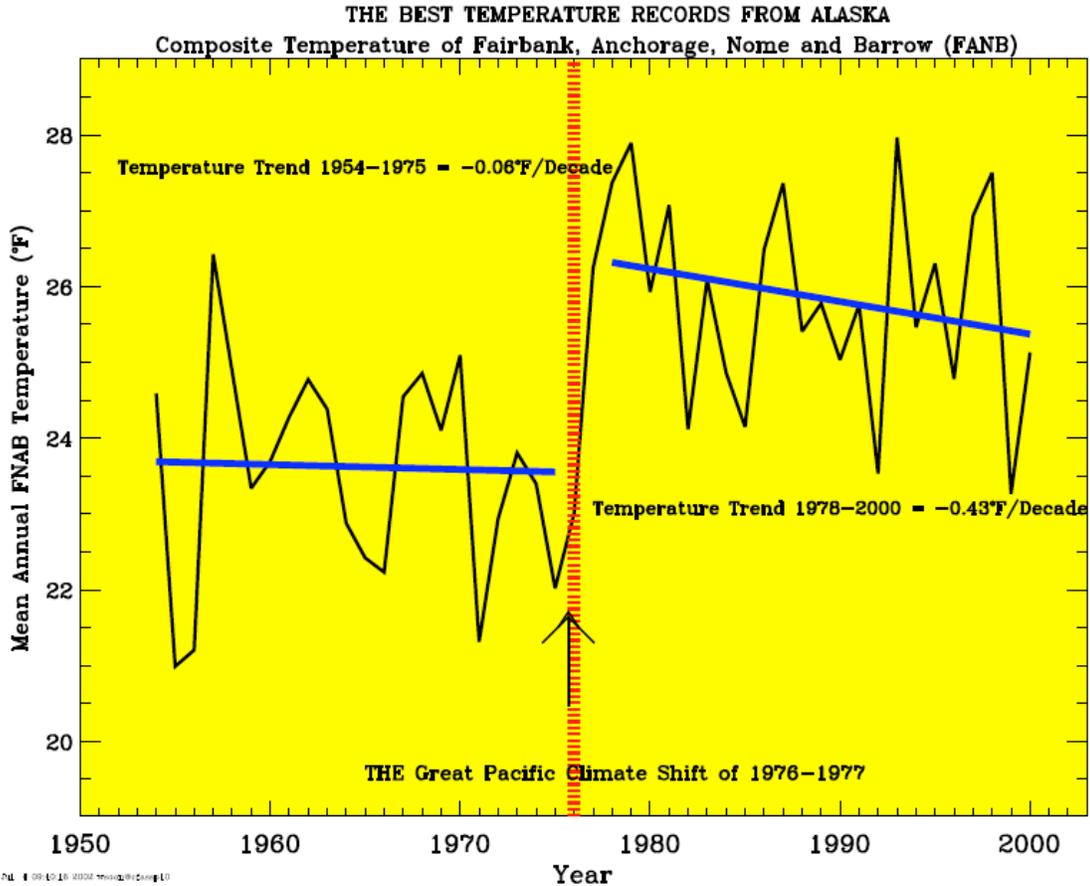
Figure 4. Composite temperature of Fairbank, Anchorage, Nome and Barrow. (Courtesy of Willie Soon.)

**SOURCES OF GLOBAL WARMING SINCE THE MID-1970s**

It is worth reiterating the observation of Visbeck, et al. that "because global average temperatures are dominated by temperature variability over the northern land masses, a significant fraction of the recent warming trend in global surface temperatures can be explained as a response to observed changes in atmospheric circulation." In addition to a positive NAO phase beginning in ~1980, and the positive PDO beginning in 1976-1977, there are also the effects of aerosols and the extraordinary solar activity in the last half of the 20$^{th}$ century to be considered.

**Aerosols**

The Arctic is purported to be the region of the earth most sensitive to radiative forcing by rising carbon dioxide concentrations. The temperature rise there is often cited, usually without consideration being given to the PDO shift in 1976-1977, as proof of the climate impact of rising anthropogenic concentrations of greenhouse gases. But other factors, even if one excludes the PDO shift, may be responsible for most if not all of the temperature rise.

Shindell and Faluvegi[12] have looked at the impact of aerosols on Arctic climate and concluded that "decreasing concentrations of sulphate aerosols and increasing concentrations of black carbon have substantially contributed to rapid Arctic warming during the past three decades." They estimate that some 45% of the warming during this period was due to this change in both types of aerosol concentrations. What this means is that rising concentrations of carbon dioxide are not responsible for almost half of arctic warming. Temperature rise comparisons for different regions of the globe are shown in Figure 5.

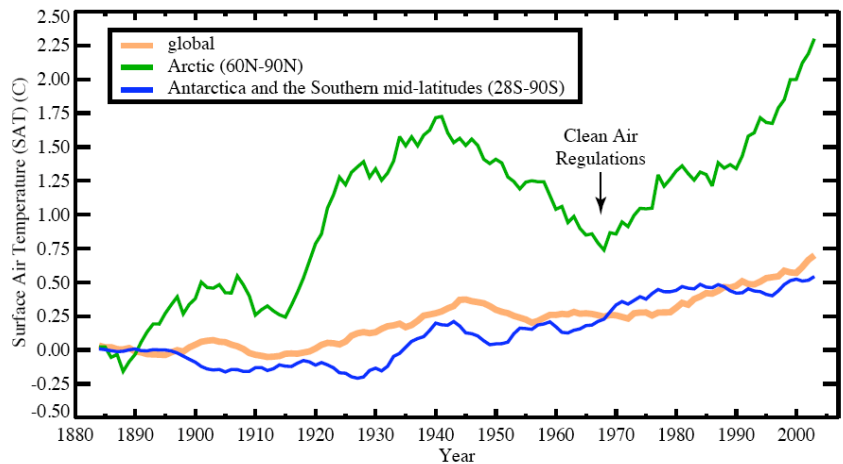

Figure 5. Relative temperature rise for different regions of the earth (from http://www.nasa.gov/topics/earth/features/warming_aerosols_prt.htm. Attributed there to Shindell.).

From Fig. 5, the temperature rise in the Arctic over the past three decades (~1978-2002) is ~1.1 °C. If 45% of this increase is due to changes in the concentrations of aerosols and black carbon, that leaves ~0.5 °C for other causes. This is obviously not compatible with the ~1.4 °C Arctic temperature rise due to the shift in the PDO in 1976-1977 shown in Fig. 4. The discrepancy may possibly be due to the use of different databases or other factors having to do with the model-based study of Shindell and Faluvegi. In any case, if the limited data in Fig. 4 is indicative of the rest of the Arctic, almost all of the Arctic warming since 1976-1977 is apparently due to causes unrelated to the rising concentrations of carbon dioxide. It would be a sophistry to claim that since the aerosols and black carbon came from burning fossil fuels there is a relationship between the carbon dioxide and the production of aerosols and black carbon—of course there is, but it is not a causal connection.

**Solar Activity**

As can be seen from Fig. 6 below,[13] the high level of solar activity during the last sixty years transcends anything seen during the last 1150 years! Notice in Fig. 6 that the variation in $^{14}C$ has an inverted scale. High solar activity, however, does not mean there are large changes in solar irradiance. That being the case, how can the small, observed variations in solar irradiance have significant impact on global temperature? One answer lies with the connection between solar activity, cosmic rays and the earth's albedo.

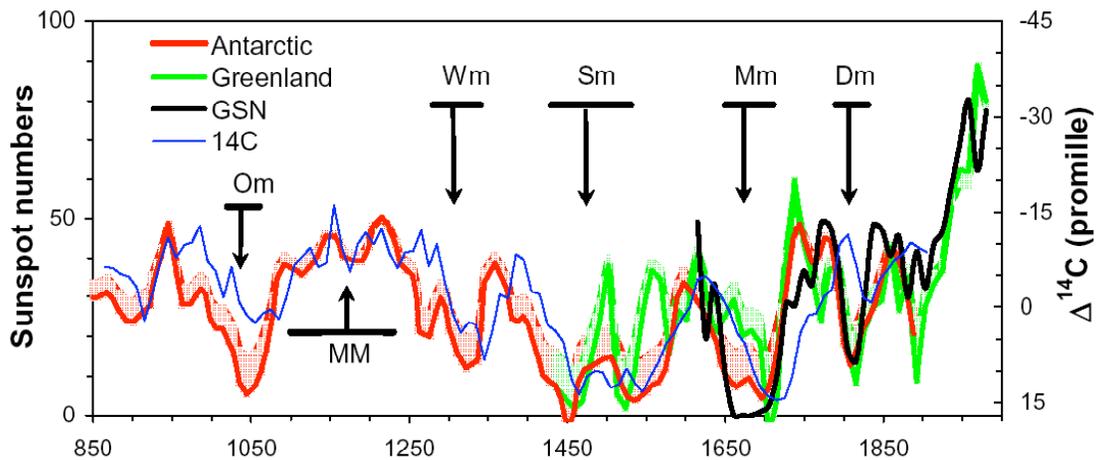

Figure 6. Time series of the sunspot number as reconstructed from $^{10}Be$ concentrations in ice cores from Antarctica (red) and Greenland (green). The corresponding profiles are bounded by the actual reconstruction results (upper envelope to shaded areas) and by the reconstructed values corrected at low values of the SN (solid curves) by taking into account the residual level of solar activity in the limit of vanishing SN. The thick black curve shows the observed group sunspot number since 1610 and the thin blue curve gives the (scaled) $^{14}C$ concentration in tree rings, corrected for the variation of the geomagnetic field. The horizontal bars with attached arrows indicate the times of great minima and maxima [J. Beer, S. Tobias, N.O. Weiss, *Solar Phys*. **181**, 237 (1998)]: Dalton minimum (Dm), Maunder minimum (Mm), Spörer minimum (Sm), Wolf minimum (Wm), Oort minimum (Om), and medieval maximum (MM). The temporal lag of $^{14}C$ with respect to the sunspot number is due to the long attenuation time for $^{14}C$ [E. Bard, G.M. Raisbeck, F. Yiou and J. Jouzel, *Earth Planet. Sci. Lett*. **150**, 453 (1997)]. Figure and modified caption from I. G Usoskin, et al., *Phys. Rev. Lett*. **91**, 211101-1 (2003). "promille" means parts per thousand.

**Variations in cloud cover and cosmic rays**

If solar variations are to play an important role in climate change, what is needed is a mechanism that transcends the effects of the relatively small variations in solar irradiance that is correlated with variations in solar activity. The most likely candidate is the modulation of the cosmic ray flux by solar activity and the observed, correlated, variations in the earth's albedo. However, cyclic variations in earth's climate following the sun's 11 year cycle cannot alone explain the warming over the last century. At most, such correlations show that solar variations can affect climate, but for solar activity to be responsible for a significant portion of the last century's warming, there must be a centennial change. And, there is. Cosmic-ray intensity, as reconstructed from $^{10}$Be concentrations in ice cores show a ~5-6% decrease over the twentieth century, corresponding to a 1% decrease in cloud cover.

The overall mechanism that has been proposed is as follows: The sun emits electromagnetic radiation and energetic particles known as the solar wind. A rise in solar activity affects the solar wind and the inter-planetary magnetic field by driving matter and magnetic flux trapped in the plasma of the local interplanetary medium outward, thereby creating what is called the heliosphere and partially shielding this volume, which includes the earth, from galactic cosmic rays—a term used to distinguish them from solar cosmic rays, which have much less energy. Long-term variations in the earth's magnetic field can also play a role. Solar variability not only affects the sun's irradiance, but also modulates incoming galactic cosmic radiation striking the earth's atmosphere. This is readily apparent in Fig. 7 below.[14] Note the inverted scale for changes in solar irradiance.

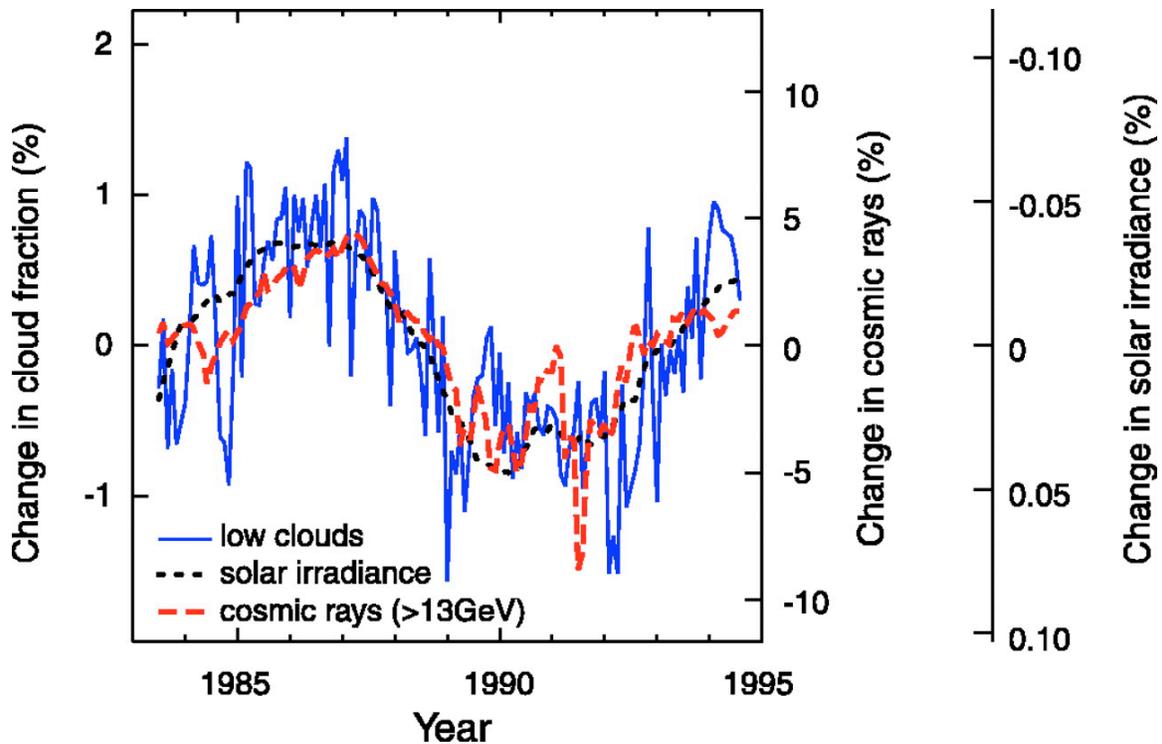

Figure 7. Variations of low-altitude cloud cover (less than about 3 km), cosmic rays, and total solar irradiance between 1984 and 1994. From K.S. Carslaw, R.G. Harrison, and J. Kirkby, *Science* **298**, 1732 (2002). Note the inverted scale for solar irradiance.

What Figure 7 shows is the very strong correlation between galactic cosmic rays, solar irradiance, and *low* cloud cover: When solar activity decreases, with a consequent small decrease in irradiance, the number of galactic cosmic rays entering the earth's atmosphere increases as does the amount of low cloud cover. This increase in cloud cover results in an increase in the earth's albedo, thereby lowering the average temperature. The sun's 11 year cycle is therefore not only associated with changes in irradiance, but also with changes in the solar wind, which in turn affect cloud cover by modulating the cosmic ray flux. This, it is argued, constitutes the strong positive feedback needed to explain the significant impact of small changes in solar activity on climate.

Finally, Figure 8 shows the centennial variation in low cloud cover derived from a variety of indices.[15]

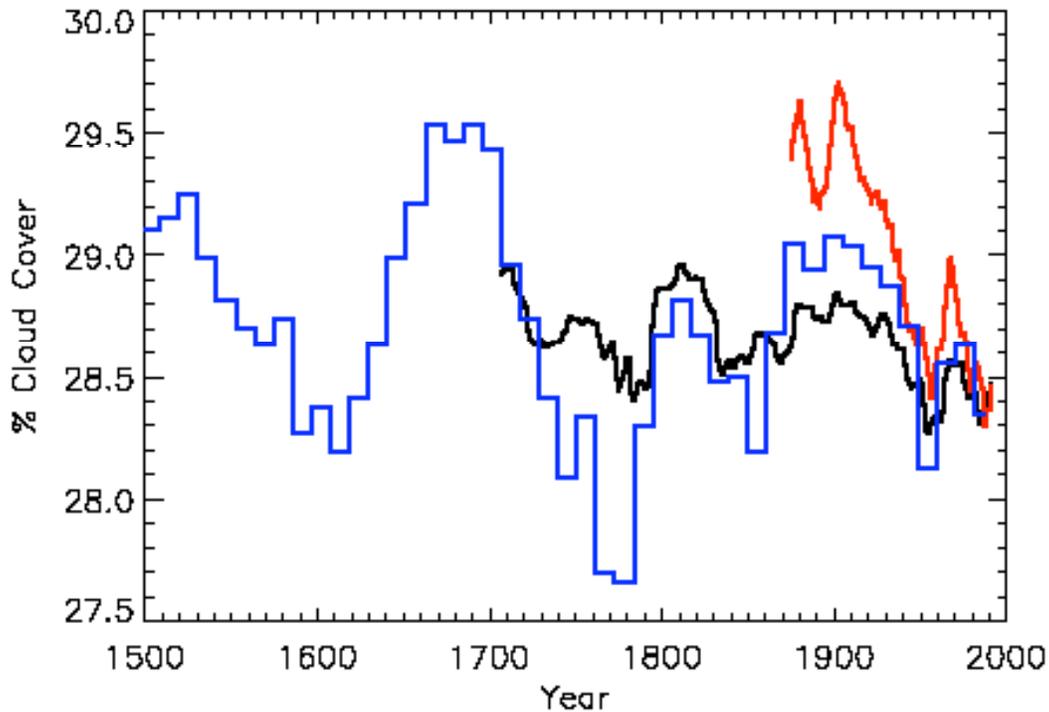

Figure 8. The 11-year smoothed reconstructed cloud cover for the whole earth derived from the Zurich Sunspot number (middle curve), the aa index (top curve), and the heliocentric potential (the curve extending from the year 1500). From E.P. Bago and C.J. Butler, *Astronomy and Geophysics* **41**, 18 (2000).

In Fig. 8, the "heliocentric potential" is an electric potential centered on the sun, which is introduced to simplify calculations by substituting electrostatic repulsion for the interaction of cosmic rays with the solar wind. Its magnitude is such that the energy lost by cosmic rays in traversing this electric field to reach the earth's orbit is equal to the energy that would be lost by cosmic rays while interacting with the solar wind in passing through the solar system to reach the earth.

As can be seen from the figure, and as was mentioned earlier, there has been a decrease in low cloud cover by about 1% over the last century. This is consistent with the data in Figure 7 where a 1% decrease in cloud cover corresponds to a 5-6% decrease in galactic cosmic ray flux. Note also, the significant rise in cloud cover during the Little Ice Age (LIA) extending from the mid 17$^{th}$ century to the early 18$^{th}$.

A valuable resource for the connection between solar activity, cosmic rays, and clouds appears in the first two sections of the CERN Cloud Proposal.[16] In the summary the authors state the following:

> "Beyond its semi-periodic 11-year cycle, the Sun displays unexplained behaviour on longer timescales. In particular, the strength of the solar wind and the magnetic flux it carries have more than doubled during the last century. The extra shielding has reduced the intensity of cosmic rays reaching the Earth's atmosphere by about 15%, globally averaged. This reduction of cosmic rays over the last century is independently indicated by the light radioisotope record in the Greenland ice cores. If the link between cosmic rays and clouds is confirmed it implies global cloud cover has decreased during the last century. *Simple estimates indicate that the consequent global warming could be comparable to that presently attributed to greenhouse gases from the burning of fossil fuels* [emphasis added].
>
> These observations suggest that solar variability may be linked to climate variability by a chain that involves the solar wind, cosmic rays and clouds. The weak link is the connection between cosmic rays and clouds. This has not been unambiguously established and, moreover, the microphysical mechanism is not understood. Cosmic rays are the dominant source of ions in the free troposphere and stratosphere and they also create free radicals. It has been proposed that ions may grow via clustering to form aerosol particles which may ultimately become cloud condensation nuclei (CCN) and thereby seed clouds. Recently a search for massive ions in the upper troposphere and lower stratosphere was started by MPIK-Heidelberg using aircraft-based ion mass spectrometers. Preliminary results indeed indicate the presence of massive positive and negative ions. In addition to their effect on aerosol formation and growth, cosmic rays may also possibly enhance the formation of ice particles in clouds."

While "the weak link is the connection between cosmic rays and clouds", this does not mean that there is not a strong correlation between the two as shown in Fig 7 and similar figures in the CLOUD proposal. What is lacking is a full understanding of the relationship between cosmic ray ionization and cloud formation (See Carslaw, et al. referenced in the caption to Fig. 7).

One can also use a simple phenomenological approach to obtain an estimate of solar variations on climate since 1900.[17] These methods yield a range of 36-50% for the percentage of temperature rise since 1900 due to the increase in solar activity.

How does this fit with Hurrell's claim that "nearly all of the cooling in the northwest Atlantic and the warming across Europe and downstream over Eurasia since the mid-1970s results from changes in the NAO"? The answer depends on whether there is a relationship between solar activity and the NAO, and this is unknown.

One thing that should be clear at this point, however, is that the recent rise in global temperature is probably not due to rising carbon dioxide concentrations as is generally assumed. Given the uncertainties outlined above, even this basic assumption behind the findings of the Intergovernmental Panel on Climate Change (IPCC) is probably incorrect. And while rising carbon dioxide concentrations are likely to be responsible for a small portion of the warming since the mid-1970s, the IPCC has been using far too high an estimate for climate sensitivity to a doubling of carbon dioxide in its projections.

It is also important to understand the uncertainties associated with such projections. Future climate projections by the IPCC will be based on coupled ocean-atmosphere climate models. These models are validated by using past data to predict present surface temperatures. There is, however, as put by Valdes, "large intermodel variability in the prediction of present-day surface temperature for atmospheric GCMs [Global Climate Models—generally using a simplified ocean treatment rather than being coupled to an ocean circulation model]. At high latitudes the differences can exceed $10^{\circ}$C. Simulations with coupled ocean-atmosphere models will almost certainly have an even wider spread of results. . . . Thus it could be said that the models and data agree to within the error bars. However, this interpretation of modeling results is controversial since a similar argument applied to future climate predictions would suggest that the predicted change in future climates in mid- and high latitudes does not exceed the modeling errors!"[18] That is, the modeling errors could well exceed the temperature changes predicted by the models. In that case, how can one argue that model projections are a sound basis for formulating public policy?

**SUMMARY**

The conclusion of this essay can be stated in a single sentence: Much, if not all, of the warming during the late 20$^{th}$ century was most likely due to natural rather than anthropogenic causes.

# REFERENCES

[1] James W. Hurrell, "Influence of variations in extratropical wintertime teleconnections on Northern Hemisphere temperature", *Geophys. Res. Lett*. **23**, 665-668 (1996).

[2] Martin H. Visbeck, Janes W. Hurrell, Lorenzo Polvani, and Heidi M. Cullen, "The North Atlantic Oscillation: Past, present, and future", *PNAS* **98**, 12876-12877 (2001).

[3] Valérie Trouet, et al., Persistent Positive North Atlantic Oscillation Mode Dominated the Medieval Climate Anomaly", *Science* **324**, 78-80 (2009).

[4] That the AMOC might have influenced the NAO is based on the model study of T. L. Delworth and R. J. Greatbatch, "Multidecadal Thermohaline Circulation Variability Driven by Atmospheric Surface Flux Forcing", *J. Clim*. **13**, 1481-1495 (2000). The model-based studies of these authors show that "the multidecadal THC [thermohaline circulation] fluctuations are driven by a spatial pattern of surface heat flux variations that bears a strong resemblance to the North Atlantic oscillation." No conclusive evidence, however, was found "that THC variability is part of a dynamically coupled mode of the atmosphere and ocean models." One should also keep in mind that at mid-lattitudes the atmosphere carries several times more heat to the North Atlantic than the ocean. See, for example, Richard Seager, "The Source of Europe's Mild Climate", *American Scientist* **94**, 334-341 (2006), and references cited therein.

[5] C. Saigne and M. Legand, "Measurements of methanesulphonic acid in Antarctic ice", *Nature* **330**, 240-242 (1987).

[6] Regine Röthlisberger and Nerilie Abram "Sea-ice proxies in Antarctic ice cores", *PAGES News* **17**, 24-26 (2009).

[7] K. R. Briffa, et al., "Low-frequency Temperature Bariations from a Northern Tree Ring Density Network", *J. Geophys. Res*. **106D3**, 2929-2941 (2001).

[8] J. P. Pinto, R. P. Turco, and O. B. Toon, *J. Geophys. Res*. **94**, 11165 (1989).

[9] R. S. J. Sparks, et al., *Volcanic Plumes* (John Whiley & Sons, Chichester 1997), Ch. 18.

[10] Thomas J. Crowley, "Causes of Climate Change Over the Past 1000 Years", *Science* **289**, 270-277 (2000).

[11] Courtesy of Willie Soon.

[12] D. Shindell and G. Faluvegi, "Climate response to regional radiative forcing during the twentieth century", *Nature Geoscience* **2**, 294-300 (2009).

[13] I. G Usoskin, et al., *Phys. Rev. Lett*. **91**, 211101-1 (2003).

[14] K.S. Carslaw, R.G. Harrison, and J. Kirkby, *Science* **298**, 1732-1737 (2002).

[15] E.P. Bago and C.J. Butler, *Astronomy and Geophysics* **41**, 18 (2000).

[16] B. Fastrup, et al. [The CLOUD Collaboration], "A Study of the Link Between Cosmic Rays and Clouds with a Cloud Chamber at the CERN PS", http://arXiv:physics/0104048v1 (2000).

[17] G. E. Marsh, "Climate Change: The Sun's Role", http://archive.org/pdf/0706.3621 (2007).

[18] P. J. Valdes, "Warm climate forcing mechanisms": B. T. Huber, K. G. Macleod, and S. L. Wing, *Warm Climates in Earth History* (Cambridge Univ. Press, Cambridge 2000), Ch. 1.